\documentclass[preprint,prc,aps]{revtex4}
\usepackage{latexsym}
\everymath={\displaystyle}
\usepackage{rotating}
\usepackage{pgf,tikz}
\usepackage{mathrsfs}
\usetikzlibrary{arrows}
\usepackage{tkz-tab}
\def\1{\mbox{l\hspace{-0.53em}1}}

\usepackage{color}
\usepackage{graphics}
\usepackage{fancybox}
\usepackage{pifont}
\usepackage{float}
\usepackage{wrapfig}
\usepackage{pifont}
\usepackage{psboxit}
\usepackage{bbm}
\usepackage{amssymb}
\usepackage{amsmath}
\usepackage{delarray}
\usepackage{rotating}

\usepackage{graphicx} 
\newcommand{\df}[2]{\ensuremath{ {\raise
1pt\hbox{$\displaystyle #1$}\over \raise -2pt \hbox{$\displaystyle
#2$}}}}
\begin{document}
\title{X(5568) as a $s u \bar d  \bar b$ tetraquark in a simple quark model}

\author{Fl. Stancu\footnote{e-mail address: fstancu@ulg.ac.be}}
\affiliation{
 University of Li\`ege, Institute of Physics B5, Sart Tilman,
B-4000 Li\`ege 1, Belgium}

\date{\today}

\begin{abstract}
The $S$-wave eigenstates of tetraquarks of type $s u \bar d  \bar b$ with J$^{P}$ = 0$^{+}$, 1$^{+}$ and 2$^{+}$
are studied within a simple quark model with 
chromomagnetic interaction and effective quark masses extracted 
from meson and baryon spectra.
It is tempting to see if this spectrum can accommodate the new narrow structure X(5568),
observed by the D\O~~Collaboration, but not confirmed  by the LHCb Collaboration.  
If it exists, such a tetraquark is a system with four different flavors
and its study can improve our understanding of multiquark systems.
The presently calculated mass of  X(5568) agrees quite well with the experimental value of 
he D\O~~Collaboration. Predictions are also made for the spectrum of the charmed partner $s u \bar d  \bar c$.
However we are aware of the 
difficulty of extracting effective quark masses, from mesons and baryons,
to be used in multiquark systems.
\end{abstract}

\maketitle
%%%%%%%%%%%%%%%%%%%%%%%%%%%%%%%%%%%%%%%%%%%%%%%%%%%%%%%%%%%%%%%%%%%%%

\section{Introduction}
The D\O~~Collaboration \cite{D0:2016mwd} has recently observed a narrow structure named $X(5568)$
in the $B_s^0\pi^\pm$ invariant mass spectrum with 5.1$\sigma$ significance
based on 10.4 fb$^{-1}$ of $p\bar p$ collisions data at $\sqrt{s}$ = 1.96 TeV.
Its measured mass and width are $M=5567.8\pm2.9(\rm stat)^{+0.9}_{-1.9}(\rm syst)$ MeV 
and $\Gamma=21.9\pm6.4(\rm stat)^{+5.0}_{-2.5}(\rm syst)$ MeV, respectively. 
Its decay into the final state $B_s^0\pi^\pm$ suggests  that $X(5568)$ 
could be a $su\bar{b}\bar{d}$ (or $sd\bar{b}\bar{u}$) 
tetraquark  with four different flavors, among which one is heavy.
The D\O~~Collaboration suggests that, with $B_s^0\pi^+$ produced in an $S$-wave,
the quantum numbers of $X(5568)$ should be $J^P = 0^+$ and that the resonance
may be the heavy analogue of the isotriplet scalar $a(980)$ with an $s$ quark
replaced by a $b$ quark.

Shortly after the D\O~~Collaboration observation a search for the claimed X(5568) was performed 
by the LHCb Collaboration in $pp$ collision data at $\sqrt{s}$ = 7 and 8 TeV, where
no significant excess was found to confirm the existence of X(5568) \cite{LHCb}. 
More data are needed to cover a larger mass range and more decay channels in order
to determine the properties of the $s u \bar d  \bar b$ tetraquark, if it exists.

An advantage of studying a tetraquark with four different flavors is that there are no annihilation
processes like in $c \bar c q \bar q$  systems ($q = u, d, s$). These are difficult to deal with theoretically.

The D\O~~Collaboration  observation immediately stimulated theoretical interest. 
So far, approaches based on QCD sum rules
\cite{Agaev:2016mjb,Wang:2016mee,Zanetti:2016wjn,Agaev:2016ijz,Chen:2016mqt,Agaev:2016urs,Dias:2016dme,Wang:2016wkj}
quark models \cite{Wang:2016tsi,Liu:2016ogz,Chen:2016npt} or
rescattering effects \cite{Liu:2016xly} have been adopted. An SU(3) classification 
has also been made \cite{He:2016yhd}.
In Refs. \cite{Agaev:2016mjb,Wang:2016mee,Zanetti:2016wjn} scalar tetraquarks were studied
while Ref. \cite{Chen:2016mqt} considered both scalar and axial  tetraquark.
Within quark models \cite{Wang:2016tsi,Liu:2016ogz} scalar, axial and tensor tetraquarks 
were analysed.

An incentive to studying tetraquarks can be found in a recent paper by
Weinberg \cite{Weinberg:2013cfa} based on large $N_c$ QCD.
It is quite natural to inquire about the existence of exotics at large
$N_c$ inasmuch as the $1/N_c$ expansion method proposed by 't Hooft \cite{'tHooft:1973jz} 
has been very successful for
ordinary hadrons.  
According to Weinberg exotic mesons consisting of
two quarks and two antiquarks are not ruled out in large $N_c$ QCD.
The real question is the decay rate of a tetraquark.
The suggestion has been followed in subsequent papers as, for example,
in Refs. \cite{Knecht:2013yqa,Lebed:2013aka,Cohen:2014via,Cohen:2014tga}.

For simplicity,  in this work,  we use the schematic model of
Refs. \cite{Hogaasen:2005jv,Buccella:2006fn,Hogaasen:2013nca} based on the chromomagnetic interaction 
between quarks (antiquarks).
In Ref. \cite{Hogaasen:2005jv}
it was shown that X(3872) can be interpreted  as a $c \bar c q \bar q$ tetraquark
where the lowest 
1$^{++}$ state has a dominant color octet-octet  component (0.9997)
and a very small color singlet-singlet   component (0.026)
which may explain why this state decays with a very small width
into $J/\psi + \rho$ or 
$J/\psi + \omega$, in agreement with the experimental
value for the total width $\Gamma <$ 2.3 MeV of X(3872)\ \cite{Choi:2003ue}.
The analysis 
has been extended to 0$^{++}$ and  2$^{++}$ sectors and the problem of
extraction the chromomagnetic strength from meson and baryons to be used 
in tetraquarks has been extensively discussed in Ref. \cite{Buccella:2006fn}.
The study of the entire spectrum of the $c \bar c s \bar s$  
system in view of the interpretation of the Y(4140) resonance as a tetraquark has been later performed in 
Ref. \cite{Stancu:2009ka} within the same model.   The suggestion was
that Y(4140) could be the strange partner of X(3872), because  
the corresponding wave function also has a very small color singlet-singlet  component
which may explain a narrow width in the $J/\psi + \phi$ channel.
The model of Refs. \cite{Hogaasen:2005jv,Buccella:2006fn} has also been considered in Ref. \cite{Cui:2006mp}.

The paper is organized as follows. In Sec. \ref{HAM}   we introduce
the quark model used in this study. In Sec. \ref{BS} we recall 
the basis states in the diquark picture,  the direct meson-meson channel useful for  $B_s^0\pi^\pm$ decays
and the exchange meson-meson channel useful for $B^+ \overline{K}^0$ decays.

In Sec.  \ref{ME}  we shortly introduce the Hamiltonian matrix of the model, followed by 
the analytic forms we have obtained  for the hyperfine interaction for $J^{P} = 0^{+}, 1^{+}$ and $2^{+}$  
states  in Secs. \ref{ZERO}, \ref{ONE} and \ref{TWO} respectively. In Sec. \ref{SPECTRUM} we exhibit 
the spectrum and the structure of the lowest eigenstates of 
the tetraquark system $su\bar{d}\bar{b}$
resulting from the diagonalization of the matrices obtained for the chromomagnetic interaction.
In Sec. \ref{sudc} we briefly present the spectrum of the $su\bar{d}\bar{c}$ tetraquark.
The last section is devoted to conclusions.

%%%%%%%%%%%%%%%%%%%%%%%%%%%%%%%%%%%%%%%%%%%%%%%%%%%%%%%%%%%%%%%%
\section{The model}\label{HAM}

This study is based
on the simple model of Refs. \cite{Hogaasen:2005jv,Buccella:2006fn,Hogaasen:2013nca}
which can reveal the  
gross features of a tetraquark, in particular the
structure of the wave functions.

In the next section we introduce the relevant 
basis states in the color-spin space, including both the color
singlet-singlet  and the octet-octet channels, the latter being called hidden color 
channels. 
There are no correlated quarks or
diquarks, as in the  diquark-antidiquark picture \cite{Maiani:2004vq,Drenska:2009cd}. 

According to Refs. \cite{Hogaasen:2005jv,Buccella:2006fn,Hogaasen:2013nca} the mass 
of a tetraquark is given by the
expectation value of the effective Hamiltonian 
\begin{equation}\label{eq:Mcm}
H =\sum_i m_i +  H_{\mathrm{CM}} ,
\end{equation}
where the chromomagnetic hyperfine interaction $H_{\mathrm{CM}}$ is described by 
\begin{equation}\label{eq:Hcm}
H_{\mathrm{CM}} = - \sum_{i,j} C_{ij}\,
~\lambda^c_{i} \cdot \lambda^c_{j}\,\vec{\sigma}_i \cdot 
\vec{\sigma_j}~.
\end{equation}
The first term in Eq. (\ref{eq:Mcm}) contains the effective quark masses $m_i$ 
as parameters. The constants $ C_{ij}$ in ($\ref{eq:Hcm}$)
represent integrals in the orbital space of some unspecified 
radial forms of the chromomagnetic part of the one gluon-exchange 
interaction potential and of the wave functions.

A warning should be given to the way of determining the effective masses $m_i$
to be used for  multiquark systems.
Besides the kinetic energy contribution, they incorporate the effect of the 
confinement, which 
is still an open problem 
\cite{Vijande:2007ix}. Information from lattice calculations on tetraquarks 
(see \emph{i. e.} \cite{Takahashi:2000te,Okiharu:2004ve})
may lead to a better understanding of the effective masses
to be used in simple models.

Presently, we  
use the compromise proposed in Refs. \cite{Hogaasen:2005jv,Buccella:2006fn} for $m_{q}$ and $m_{s}$,
while for $m_{b}$ we rely on a value compatible with heavy-light systems \cite{Hogaasen:2013nca}. Therefore
we have
\begin{equation}\label{eq:mass}
\begin{array}{lll}
m_{u,d}=320\,\mathrm{MeV}, &
m_{s}=590\,\mathrm{MeV},&
 m_{b}=4860\,\mathrm{MeV}.
%m_{s}=590\,\mathrm{MeV},&
%m_{q}=320\,\mathrm{MeV}.%\\
\end{array}\end{equation}
Due to the arbitrariness in the choice of effective masses of quarks, 
precise estimates of the absolute values of tetraquark masses 
is difficult to make. One can have an approximate idea about the range 
where the spectrum should be located. 
We stress that in this study we favor $m_{u,d}$ = 320 MeV instead of 
450 MeV of Ref. \cite{Hogaasen:2013nca} because this value brings 
the $\rho$ and the $\pi$ masses close to experiment (see below).

However,  the relative distances between the eigenstates obtained  
from the chromomagnetic Hamiltonian  (\ref{eq:Hcm}) and the structure 
of its eigenstates do not depend on the effective masses,
which is important for exploring the strong decay properties.

The  parameters $C_{ij}$ have  been taken from  
Ref.\  \cite{Buccella:2006fn}, 
Table 1 for $C_{q \bar q}$  and Table 2 for $C_{qq}$ ($q = u,d$). The parameter
 $C_{\bar d \bar b}$, absent in Ref.\  \cite{Buccella:2006fn}, was taken identical to $C_{db}$ of Ref. \cite{Hogaasen:2013nca}.
Therefore we have
  
\begin{equation}\label{eq:par}
\begin{array}{lll}
  C_{u \bar d}=29.8\,\mathrm{MeV},&
C_{s \bar d}=18.4\,\mathrm{MeV},&C_{us}=13\,\mathrm{MeV},\\
%
%C_{ss}=10\,\mathrm{MeV},&
C_{u \bar b}=2.1\,\mathrm{MeV}, &C_{\bar d \bar b}=1.9\,\mathrm{MeV},
%C_{c \bar s}=2.1\,\mathrm{MeV}, 
&C_{s \bar b} = 2.2\,\mathrm{MeV}. 
\end{array}\end{equation}
We should mention that the above parameters were extracted from a
global fit to meson and baryon ground states. With these parameters 
the pion acquires a mass of 163 MeV and the $\rho$ meson 799 MeV,
the experimental values being approximately 140 MeV and 770 MeV respectively.

As one can infer from
these values, most of the hyperfine attraction will come from light
$q q$ or $q \bar q$ pairs, as expected. 
%%%%%%%%%%%%%%%%%%%%%%%%%%%%%%%%%%%%%%%%%%%%%%%
% May 4, 2016

In Ref. \cite{Liu:2016ogz} the same type 
of Hamiltonian has been used with hyperfine interaction  parameters $C_{ij}$ 
obtained from a fit to light or
heavy baryons and $B$ and $D$ mesons. 
The comparison of the values of the relevant common parameters of Ref. \cite{Liu:2016ogz}  
with those of Ref. \cite{Buccella:2006fn} shows that they are very close
to each other, except for $C_{us}$ which is 8.8 MeV in Table VI Ref. \cite{Liu:2016ogz}
(for consistency with
the present definition the values of Table VI Ref. \cite{Liu:2016ogz}
should be multiplied by a factor 3/32).  In fact $C_{us}$ = 13 MeV of 
Eq. (\ref{eq:par}) is the central value of this parameter varying between 
12 MeV and 14 MeV in  Ref. \cite{Buccella:2006fn}.
Taking $C_{us}$ = 8.8 MeV (which fits well the $\Sigma^*$ - $\Sigma$ splitting \cite{Liu:2016ogz}),
instead of 13 MeV, increases the mass of the lowest state
by only 10 MeV.

However the most important parameter (see next sections),
namely $C_{u \bar d}$, is missing in  Ref. \cite{Liu:2016ogz}. The lack of $C_{u \bar d}$
from  Table VI Ref. \cite{Liu:2016ogz} was justified  by the fact that  pseudoscalar
mesons are influenced by chiral symmetry and its spontaneous breaking.
By using a triquark-heavy antiquark basis states the masses of tetraquarks were 
calculated by avoiding this parameter, as any other $C_{ij}$ parameter of
the chromomagnetic 
interaction between quarks and antiquarks, such contribution being claimed to vanish 
in the most significant case.
In addition, the mixing of the four basis states constructed from the 
triquark-heavy antiquark basis was neglected which made an extra loss 
in the contribution of the hyperfine  interaction.
In our case  $C_{u \bar d}$ of Eq. (\ref{eq:par})
gives a $\rho -\pi$ splitting of 635.7 Mev as compared to the experimental value of 630 MeV.

%%%%%%%%%%%%%%%%%%%%%%%%%%%%%%%%%%%%%%%%%%%%%%%%%%%%%%%%%%%%%%%%%%

We do not intend to make a fine tuning of the effective masses. 
We are mostly interested 
in the structure of the tetraquark wave functions which essentially depends 
on the hyperfine interaction as well as on the level sequence.
We shall compare the calculated spectrum to the experimental thresholds.

%%%%%%%%%%%%%%%%%%%%%%%%%%%%%%%%%%%%%%%%%%%%%%%%%%%%%%%%%%%%%%%%%
\section{The basis states}\label{BS}

Here we use a basis vectors relevant for understanding the decay properties
of tetraquarks.  The total wave function of a tetraquark is a linear 
combination of these basis vectors.
We suppose that particles 1 and 2 are quarks and particles 3 and 4 are
antiquarks, see Fig. 1. Presently we take 1 = $u$, 2 = $s$, 3 = $\overline d$ and 4 = $\overline b$.

 In principle the basis vectors should contain
the orbital, color, flavor and spin degrees of freedom such as to account 
for the Pauli principle. But, as we consider $\ell = 0$ states 
the orbital part is symmetric and anyhow irrelevant for the effective
Hamiltonian described in the previous section. Moreover, 
as the flavor operators do not explicitly appear in
the Hamiltonian, the flavor part does not need to be specified. 
A detailed description of each of the three distinct bases  corresponding to 
three choices of internal coordinates, shown in Fig. 1, was presented in 
Refs.\ \cite{Brink:1994ic,Brink:1998as}. One can use any base, $(a)$, $(b)$
or $(c)$.

%%%%%%%%%%%%%%%%%%%%%%%%%%%%%%%%%%%%%%%%%%%%%%%%%%%%%%%%%%ù
%
%  From arno2.tex 
%
%%%%%%%%%%%%%%%%%%%%%%%%%%%%%%%%%%%%%%%%%%%%%%%%%%%%%%
%    Before May 14, 2016
%
%\begin{figure}\label{fig1}
%\begin{center}
%\includegraphics*[width=10.0cm,keepaspectratio]{coord150.eps}
%\end{center} 
%%%%%%%%%%%%%%%%%%%%%%%%%%%%%%%%%%%%%%%%%%%
\begin{figure}\label{fig1}
\begin{center}
\resizebox{0.3 \linewidth}{0.40 \linewidth}{%
\begin{tikzpicture}[line cap=round,line join=round,>=triangle 45,x=1.0cm,y=1.0cm]
\clip(-1,-4.5) rectangle (5,3.5); %size of the figure: from (-1,-3) to (5,3)
\draw [->,line width=2.4pt] (0.,-2.) -- (0.,2.); %draw vector from (0.,-2.) to (0.,2.)
\draw [->,line width=2.4pt] (0.,0.) -- (2.5,-0.5);
\draw [->,line width=2.4pt] (1.5,-2.) -- (3.76,1.52);
\begin{scriptsize}
\draw [fill=black, line width=1mm] (0.,2.2) circle (6pt); %draw a filled circle of line width=1mm and radius 6pt at (0.,2.2)
\draw[color=black] (-0.02,2.8) node {\huge 1}; %write "1" at (-0.02,2.8)
\draw [fill=black, line width=1mm] (0.,-2.2) circle (6pt);
\draw[color=black] (-0.02,-2.8) node {\huge 2};
%\draw[color=black] (1.,0.4) node {\huge$\vec{\lambda}$};
\draw [color=black, line width=1mm] (3.8,1.68) circle (6pt);
\draw[color=black] (3.92,2.24) node {\huge 3};
\draw [color=black, line width=1mm] (1.38,-2.18) circle (6pt); %draw an empty circle of line width=1mm and radius 6pt at (0.,2.2)
\draw[color=black] (1.36,-2.8) node {\huge 4};

\draw[color=black] (2,-4) node {\large (a)};
\end{scriptsize}
\end{tikzpicture}
}
$\ $
\resizebox{0.3 \linewidth}{0.4 \linewidth}{%
\begin{tikzpicture}[line cap=round,line join=round,>=triangle 45,x=1.0cm,y=1.0cm]
\clip(-2,-4.5) rectangle (5,3);
\draw [->,line width=2.4pt] (1.5,-2) -- (-1.5,-2);
\draw [->,line width=2.4pt] (0,-2) -- (1.5,0);
\draw [->,line width=2.4pt] (4,-0.7) -- (-0.8,0.7);
\begin{scriptsize}
\draw [fill=black, line width=1mm] (-1,0.8) circle (6pt);
\draw[color=black] (-1,1.4) node {\huge 1};
\draw [fill=black, line width=1mm] ((-1.7,-2) circle (6pt);
\draw[color=black] ((-1.7,-2.6) node {\huge 2};
%\draw[color=black] (1.4,-1) node {\huge$\vec{x}$};
\draw [color=black, line width=1mm] (4.25,-0.8) circle (6pt);
\draw[color=black] (4.25,-0.2) node {\huge 3};
\draw [color=black, line width=1mm] (1.7,-2) circle (6pt);
\draw[color=black] (1.7,-2.6) node {\huge 4};

\draw[color=black] (1.5,-4) node {\large (b)};
\end{scriptsize}
\end{tikzpicture}
}
$\ $
\resizebox{0.3 \linewidth}{0.4 \linewidth}{%
\begin{tikzpicture}[line cap=round,line join=round,>=triangle 45,x=1.0cm,y=1.0cm]
\clip(-2,-4.5) rectangle (5,3);
\draw [->,line width=2.4pt] (1.55,-1.8) -- (-0.8,0.7);
\draw [->,line width=2.4pt] (2.9,-1) -- (0.2,-0.4);
\draw [->,line width=2.4pt] (4,-0.8) -- (-1.5,-2);
\begin{scriptsize}
\draw [fill=black, line width=1mm] (-1,0.8) circle (6pt);
\draw[color=black] (-1,1.4) node {\huge 1};
\draw [fill=black, line width=1mm] ((-1.7,-2) circle (6pt);
\draw[color=black] ((-1.7,-2.6) node {\huge 2};
%\draw[color=black] (1.4,0) node {\huge$\vec{y}$};
\draw [color=black, line width=1mm] (4.25,-0.8) circle (6pt);
\draw[color=black] (4.25,-0.2) node {\huge 3};
\draw [color=black, line width=1mm] (1.7,-2) circle (6pt);
\draw[color=black] (1.7,-2.6) node {\huge 4};

\draw[color=black] (1.5,-4) node {\large(c)};
\end{scriptsize}
\end{tikzpicture}
}
\end{center}

\caption{Three independent relative coordinate systems. Solid and 
open circles represent quarks and antiquarks respectively: (a)
diquark-antidiquark channel, (b) direct meson-meson channel, (c) exchange
meson-meson channel. }
\end{figure}
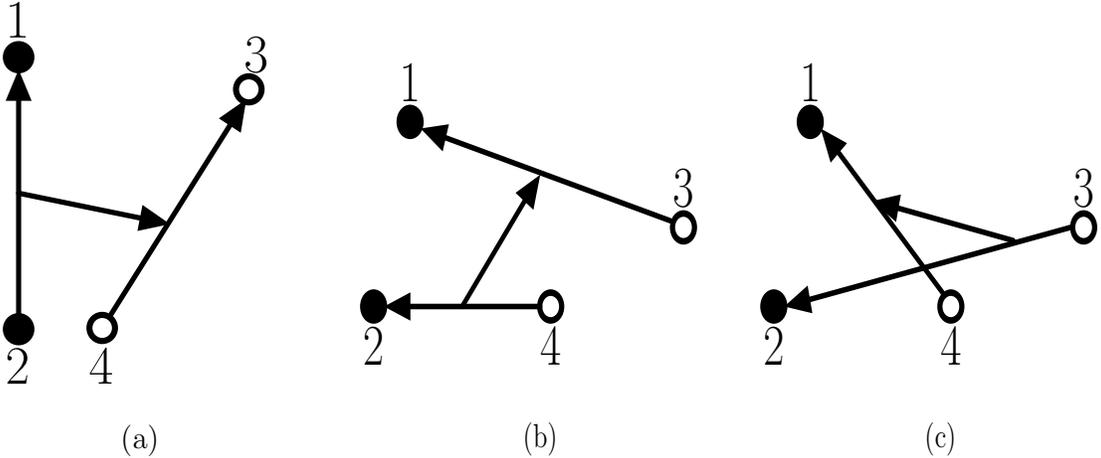
%%%%%%%%%%%%%%%%%%%%%%%%%%%%%%%%%%%%%%%%%%%%%%%%%%%%%%%%

In the color space the three 
distinct bases are: 
%\begin{equation}\label{diquark}
~$ a)~|\overline{3}_{12} 3_{34} \rangle, ~ |6_{12} \overline{6}_{34} \rangle $,
%\end{equation}
%\begin{equation}\label{directcolor}
~$ b)~|1_{13} 1_{24} \rangle, ~ |8_{13} 8_{24} \rangle $ ,
%\end{equation} 
and 
%\begin{equation}\label{exchangecolor}
~$ c)~|1_{14} 1_{23} \rangle, ~ |8_{14} 8_{23} \rangle$, 
%\end{equation} 
associated to the three distinct internal coordinate systems 
shown in Fig. 1.
The 3 and $\overline{3}$ are antisymmetric and 6 and $\overline{6}$
are symmetric under interchange of quarks and antiquarks respectively.	
This basis is convenient for diquark-antidiquark models, 
where the color space is truncated to contain only 
$|\overline{3}_{12} 3_{34} \rangle$ states \cite{Maiani:2004vq,Drenska:2009cd}.
This reduces 
each J$^{PC}$ spectrum  to twice less states than allowed 
by the Pauli principle \cite{Stancu:2006st}
and influences the tetraquark properties.
In the present context an example is Ref. \cite{Wang:2016tsi} where,
although mixing across the basis states was allowed, only the color state
$|\overline{3}_{12} 3_{34} \rangle$ was considered which may partly 
explain why the contribution of the hyperfine interaction is so small.

The sets $b)$  and $c)$ both
contain a color singlet-singlet 
and a color octet-octet  state. The amplitude of the latter
vanishes asymptotically, when the mesons, into which a tetraquark 
decays, separate. These
are called \emph{hidden color} states by analogy to states
which appear in  the nucleon-nucleon problem, defined as a six-quark system
\ \cite{HARVEY,Pepin:2001is}.
The contribution of hidden color states to the 
binding energy of light tetraquarks has been calculated explicitly in 
Ref.  \cite{Brink:1994ic}.
The coordinate sets $b)$  and $c)$ define the direct and the exchange 
meson-meson channels. 
The relation between the three different bases can be found in 
Ref. \ \cite{Brink:1998as}.

As the quarks and antiquarks are spin 1/2 particles the total 
spin of a tetraquark can be $S = 0$ , $S = 1$  
or $S = 2$.
In the following we shall use the notation $P$ 
and $V$ for pseudoscalar and vector 
subsystems respectively.  

%%%%%%%%%%%%%%%%%%%%%%%%%%%%%%%%%%%%%%%%%%%%%%%%%%%%%%%%%%%%%%%%
\section{Matrix elements}\label{ME}

As already stressed, 
the Hamiltonian  (\ref{eq:Mcm}) does not contain flavor operators
so that the flavor part of the wave function does not need to be specified. 
Then the quantum numbers of the states can be defined in terms of the
permutation properties of the spin and color parts of the basis vectors, as shown in Ref. 
\cite{Stancu:2008zh}. 
The basis vectors can be written such as to have a good charge conjugation quantum number 
(see  Appendix \ref{chargeconj}).
In particular, for two identical flavors, a ground state tetraquark can have
$J^{PC} = 0^{++}, 1^{++}, 1^{+-}$ and $2^{++}$. For four distinct flavors the
situation is more complicated, as it will be seen below.

In the direct meson-meson channel, in  each case a basis can be built
with the quark-antiquark pairs (1,3) and (2,4) as subsystems, where each
subsystem has a well defined color state, a singlet-singlet or an octet-octet. 
The other quark-antiquark pairs,
(1,4) and (2,3) are needed to study the meson-meson exchange channels (see Fig. 1c).
One can fix a basis in terms of the problem one looks at, but
for convenience, in the calculations one can
pass from one basis to another by an orthogonal transformation.

%%%%%%%%%%%%%%%%%%%%%%%%%%%%%%%%%%%%%%%%%%%%%%%%%%%%%%%%%%%%%%%%%%%%%%%%%
\section{The J$^{P}$ = 0$^{+}$ states}\label{ZERO}

To study the  $J^{P} = 0^{+}$ spectrum we can use %the$J^{PC} = 0^{++}$ 
a basis constructed from products of color and spin states 
associated to Fig. 1b 
\begin{eqnarray}\label{eq:betai}
\hskip -10pt& \psi^1_{0^{+}} = 
| 1_{13} 1_{24} P_{13} P_{24} \rangle,\
&  \psi^2_{0^{+}} =
| 1_{13} 1_{24} (V_{13} V_{24})_0 \rangle, \nonumber\\
\hskip -10pt& \psi^3_{0^{+}} =
| 8_{13} 8_{24} P_{13} P_{24} \rangle,\ 
&\psi^4_{0^{+}} =
| 8_{13} 8_{24}  (V_{13} V_{24})_0 \rangle .
\end{eqnarray}

The chromomagnetic interaction Hamiltonian with minus sign, -$H_{\mathrm{CM}}$,
acting on this basis leads to the following symmetric matrix
\\ 
\vskip -20pt

%\begin{table}
%\caption{The matrix for $J^{PC}$ = $0^{++}$}
%%%%%%%%%%
\begin{equation}
\begingroup
\makeatletter
\def\f@size{7}
%\checkmathfonts
\left[%\vbox{\hskip -10pt
\renewcommand{\arraystretch}{1.2}
\begin{tabular}{cccccc}
%
% row 1
$16(C_{13}+C_{24})$&0&0&$ -4 \sqrt{\df{2}{3}}(C_{12}+C_{23}+C_{14}+C_{34})$\\
%
% row 2
 &$-\df{16}{3}(C_{13}+C_{24})$&$ ~~-4 \sqrt{\df{2}{3}}(C_{12}+C_{23}+C_{14}+C_{34})$
&$\df{8 \sqrt2}{3}(C_{23}-C_{12}+ C_{14}-C_{34})$\\
%
% row 3
%
 & &$-2(C_{13}+C_{24})$ & $\df{2}{\sqrt3}[2(C_{12}+C_{34})-7(C_{23}+C_{14})]$\\
%
% row 4
%
 & & & $\df{8}{3}(C_{12}+C_{34})+\df{28}{3}(C_{14}+C_{23})+\df{2}{3}(C_{13}+C_{24})$
\\
\end{tabular}\right]
\label{0++}
\endgroup
\end{equation}
%\end{table}
%\end{widetext}\end{sidewaystable}

%%%%%%%%%%%%%%%%%%%%%%%%%%%%%%%%%%%%%%%%%%%%%%%%%%%%%%%%%%%%%%%%%%
In the case of two distinct flavors, for
example $s \bar s c \bar c$ systems, the following equalities hold 
\begin{equation}\label{twoflavor} 
C_{14} = C_{23}, ~~~C_{12} = C_{34},
\end{equation}
in which case the above matrix takes a simpler form, to be found in Ref. \cite{Stancu:2009ka}.
Note that the matrix element row 1 column 4 of -$H_{\mathrm{CM}}$ has a correct - sign, 
a misprint has to be corrected in  Ref. \cite{Stancu:2009ka}.

%%%%%%%%%%%%%%%%%%%%%%%%%%%%%%%%%%%%%%%%%%%%%%%%%%%%%%%%%%%%%%%%%
%
% May 10, 2016
%
%%%%%%%%%%%%%%%%%%%%%%%%%%%%%%%%%%%%%%%%%%%%%%%%%%%%%%%%%%%%%%%%%

\section{The J$^{P}$ = 1$^{+}$ states}\label{ONE}

To study the  $J^{P} = 1^{+}$ spectrum we find useful to start from a
basis constructed from products of color 
and spin states associated to Fig. 1a (diquark basis). These are

\begin{eqnarray}\label{eq:psi}
\hskip -10pt& \psi^1_{1^{+}} = 
%(q_1 \overline{q}_3)^1_0 \otimes (q_2 \overline{q}_4 )^{{1}}_{0}~,\
| 6_{12} \bar 6_{34} (V_{12} V_{34})_1 \rangle,\
&  \psi^2_{1^{+}} =
%(q_1 \overline{q}_3)^1_1 \otimes (q_2 \overline{q}_4)^{{1}}_{1},\nonumber\\
| \bar 3_{12} 3_{34} (P_{12} V_{34})_1 \rangle, \nonumber\\
\hskip -10pt& \psi^3_{1^{+}} =
%(q_1 \overline{q}_3)^8_0 \otimes (q_2\overline{q}_4 )^{8}_{0},\
| \bar 3_{12} 3_{34} (P_{12} V_{34})_1 \rangle,\ 
&\psi^4_{0^{+}} =
%(q_1 \overline{q}_3)^8_1 \otimes (q_2 \overline{q}_4)^{{8}}_{1}, \\
| 6_{12} \bar 6_{34}  (V_{12} P_{34})_1 \rangle, \nonumber\\
\hskip -10pt& \psi^5_{1^{+}} =| \bar 3_{12} 3_{34} (V_{12} P_{34})_1 \rangle,\ 
&\psi^6_{0^{+}} =| 6_{12} \bar 6_{34}  (P_{12} V_{34})_1 \rangle.
\end{eqnarray}
Using this basis we now construct a new basis where every vector has a definite charge
conjugation, following the observation made in the Appendix. They are defined as
the linear combinations 
\begin{eqnarray}\label{eq:cargec}
\hskip -10pt& \psi^1_{1^{+-}} = \psi^1_{1^{+}}, 
&             \psi^2_{1^{+-}} = \psi^2_{1^{+}},  \nonumber\\
\hskip -10pt& \psi^3_{1^{+-}} = \frac{1}{\sqrt{2}}( \psi^3_{1^{+}} - \psi^5_{1^{+}}),
&             \psi^4_{1^{+-}} = \frac{1}{\sqrt{2}}( \psi^4_{1^{+}} - \psi^6_{1^{+}}),\nonumber\\
\hskip -10pt& \psi^5_{1^{++}} = \frac{1}{\sqrt{2}}( \psi^3_{1^{+}} + \psi^5_{1^{+}}),
&             \psi^6_{1^{++}} = \frac{1}{\sqrt{2}}( \psi^4_{1^{+}} + \psi^6_{1^{+}}).
\end{eqnarray}
In this basis the matrix of -$H_{\mathrm{CM}}$ for $J^+$ = $1^+$ has the form
\begin{equation}
\begingroup
\makeatletter
\def\f@size{1}
%\checkmathfonts
\left[%\vbox{\hskip -10pt
\renewcommand{\arraystretch}{1.2}
\begin{tabular}{cccccc}
%
% row 1
$D_{11}$ & ~~ $2\sqrt{2}(C_{13}+C_{24}-C_{14}-C_{23})$ & $ ~~ 4\sqrt{2}(C_{13}-C_{24})$ 
&~~ -$\frac{20}{3}(C_{13}-C_{24})$ & $-4\sqrt{2}(C_{14}-C_{23})$ &$ \frac{20}{3}(C_{14}-C_{23})$\\
%
% row 2
 & $D_{22}$ & $\df{8}{3}(C_{13}-C_{24})$ & -$4\sqrt{2}(C_{13}-C_{24})$ &
 $\frac{8}{3}(C_{14}-C_{23})$ &    $- 4\sqrt{2}(C_{14}-C_{23})$   \\
%
% row 3
%
 & & $D_{33}$ & $-2\sqrt{2}(C_{13}+C_{14}+C_{23}+C_{24})$ & $\frac{16}{3}(C_{12}-C_{34})$ & 0 \\
%
% row 4
%
 & & &  $D_{44}$   &     0    & $ \frac{8}{3}(C_{12}-C_{34})$ \\
%
% row 5 
 & & & & $D_{55}$  & $ -2 \sqrt{2}(C_{13}+C_{14}+C_{23}+C_{24}) $ \\
%
% row 6
 & & & & & $D_{66}$ \\
\end{tabular}\right]
\label{1++}
\endgroup
\end{equation}
%\end{table}
%\end{widetext}\end{sidewaystable}
with 
\begin{eqnarray}
D_{11} =   \frac{4}{3}(C_{12}+C_{34})+\frac{10}{3}(C_{13}+C_{14}+C_{23}+C_{24}) \nonumber \\
D_{22} = - \frac{8}{3}(C_{12}+C_{34})+\frac{4}{3}(C_{13}+C_{14}+C_{23}+C_{24}) \nonumber \\
D_{33} =   \frac{8}{3}(C_{12}+C_{34})+\frac{4}{3}(C_{13}-C_{14}-C_{23}+C_{24})  \nonumber \\
D_{44} = - \frac{4}{3}(C_{12}+C_{34})+\frac{10}{3}(C_{13}-C_{14}-C_{23}+C_{24})  \nonumber \\
D_{55} =   \frac{8}{3}(C_{12}+C_{34})-\frac{4}{3}(C_{13}-C_{14}-C_{23}+C_{24}) \nonumber \\
D_{66} = - \frac{4}{3}(C_{12}+C_{34})-\frac{10}{3}(C_{13}-C_{14}-C_{23}+C_{24})
\end{eqnarray}
One can see again that for tetraquarks with two flavors, where the relations (\ref{twoflavor})
remain valid, the above matrix takes a quasidiagonal form, with one block 4 $\times$ 4 for states 
with $ C $ = - 1 and a 2 $\times$ 2 block for states with  $ C $ = + 1 respectively.  For 
$s \bar s c \bar c$ systems, for example, their eigenvalues recover the $1^{+-}$ and  $1^{++}$
spectra of Fig. 2 Ref. \cite{Stancu:2009ka}, obtained here in another basis. The advantage of
using the basis (\ref{eq:cargec}) is that one can have a control of the wave function components
with a specific charge conjugation and this may be useful in processes where the charge conjugation
quantum number is conserved.  

%%%%%%%%%%%%%%%%%%%%%%%%%%%%%%%%%%%%%%%%%%%%%%%%%%%%%%%%%%%%%%%%%%%%%%%%%%%%%%%

\section{The J$^{P}$ = 2$^{+}$ states}\label{TWO}

For tensor tetraquark one can use  a basis of color and  spin states 
corresponding to Fig. 1b. This is 
 \begin{eqnarray}\label{eq:tensor}
\hskip -10pt& \psi^1_{2^{++}} = 
%(q_1 \overline{q}_3)^1_0 \otimes (q_2 \overline{q}_4 )^{{1}}_{0}~,\
| 1_{13}  1_{24} (V_{13} V_{24})_2 \rangle,\
&  \psi^2_{2^{++}} =
%(q_1 \overline{q}_3)^1_1 \otimes (q_2 \overline{q}_4)^{{1}}_{1},\nonumber\\
| \bar 8_{13} 2_{24} (V_{13} V_{24})_2 \rangle.
\end{eqnarray}
The corresponding  ~-$H_{\mathrm{CM}}$ 2$\times$2 matrix is
\\
\vskip -20pt
\begin{equation}\left[%\vbox{\hskip -4pt
\renewcommand{\arraystretch}{2.0}
\begin {tabular}{cc}
%
% row 1
$ -\df{16}{3}(C_{13}+C_{24})$ & $ {\df{4 \sqrt2}{3}}(C_{12}+C_{34}-C_{14}-C_{23})$\\
%
% row 2
 & $ - \df{2}{3} (2C_{12}+2C_{34}+7C_{23}+7C_{14}-C_{13}-C_{24})$ \\
\end{tabular} \right]
\label{2++}
\end{equation}
where one can use the relations (\ref{twoflavor}) to recover the result for
two flavors of  Ref. \cite{Stancu:2009ka}.

%%%%%%%%%%%%%%%%%%%%%%%%%%%%%%%%%%%%%%%%%%%%%%%%%%%%%%%%%%%%%%%%%%%%%%%%%%%%%
We  note that
the above matrices can be used in any quark model containing a chromomagnetic interaction.
In that case the parameters $C_{ij}$ should be replaced by integrals
containing the chosen form factor of the chromomagnetic interaction and the 
orbital wave functions of the model.

The matrices (\ref{0++}),  (\ref{1++}) and   (\ref{2++}) can be used to
calculate the spectrum of either  $s u \bar d  \bar b$ or its charm partner $s u \bar d  \bar c$ 
by implementing the corresponding parameters $C_{ij}$. Below we shall show results
for   $s u \bar d  \bar b$ 
and in the next section we shall shortly describe the spectrum of $s u \bar d  \bar c$ .

%%%%%%%%%%%%%%%%%%%%%%%%%%%%%%%%%%%%%%%%%%%%%%%%%%%%%%%%%%%%%%%

\section{The spectrum of $s u \bar d  \bar b$}\label{SPECTRUM}

%%%%%%%%%%%%%%%%%%%%%%%%%%%%%%%%%%%%%%%%%%%%%%%%%%%%%
%  May 18, 2016
%  From arno2.tex 
%%%%%%%%%%%%%%%%%%
%\vspace{2.5cm}
\begin{figure}\label{fig2}
%%%%%%%%%%%%%%%%%%%%%%%%%%%%%%%%%%%%%%%%%%%%%%%%%%%%%
%\thispagestyle{empty}
\definecolor{red}{rgb}{1,0,0} 
\begin{center}
\setlength{\unitlength}{0.75cm}
\begin{picture}(24,20)(0,0)
\def\level{\line(2,0){2}}
\thicklines
%%%%%%%%%%%%%%%%%%%%%%%%%%%%0+
\put(2.5,2.0){\large 0$^{+}$}
\put(1.5,5.530){\level}
\put(3.7,5.330){\large 5530}
 
\put(1.5,11.47){\level}
\put(3.7,11.27){\large 5827}
 
\put(1.5,17.57){\level}
\put(3.7,17.37){\large 6132}
 
\put(1.5,21.75){\level}
\put(3.7,21.55){\large 6341}
%%%%%%%%%%%%%%%%%%%%%%%%%%%%%1+
\put(6.7,2.0){\large  1$^{+}$}

\put(6.0,6.69){\level}
\put(8.2,6.59){\large 5588}

\put(6.0,12.09){\level}
\put(8.2,11.89){\large 5858}

\put(6.0,17.40){\level}
\put(8.2,17.20){\large 6123}

\put(6.0,18.10){\level}
\put(8.2,17.90){\large 6158}

\put(6.0,19.63){\level}
\put(8.2,19.43){\large 6235}

\put(6.0,21.19){\level}
\put(8.2,20.99){\large 6313}

%%%%%%%%%%%%%%%%%%%%%%%%%2+
\put(12.0,2.0){\large  2$^{+}$}

\put(11.0,18.59){\level}
\put(13.2,18.39){\large 6183}

\put(11.0,20.17){\level}
\put(13.2,19.97){\large 6262}
%%%%%%%%%%%%%%%%%%%%%%%%%
\end{picture}
\end{center}
\caption{The spectrum of the $s u \bar d  \bar b$ tetraquark. } 
\end{figure}
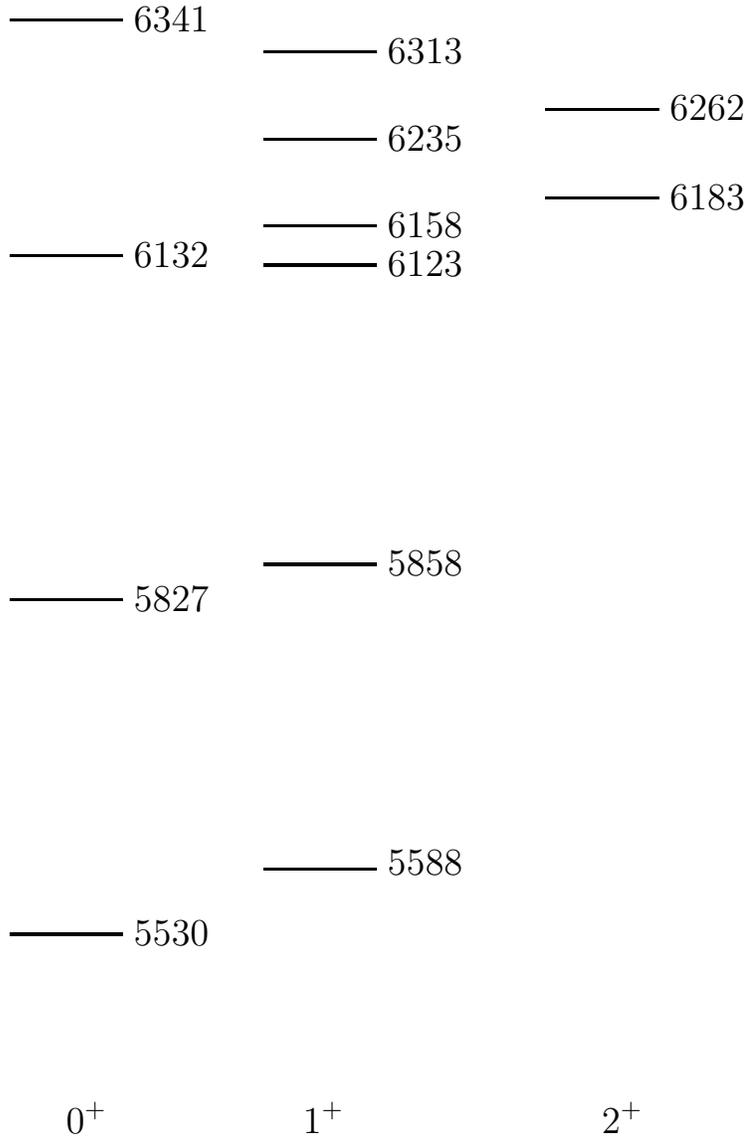

%%%%%%%%%%%%%%%%%%%%%%%%%%%%%%%%%%%%%%%%%%%%%%%%

The calculated spectrum of the  $s u \bar d  \bar b$ tetraquark is exhibited in Fig. 2.
One can see that
the choice of the effective masses (\ref{eq:mass}) and of the  hyperfine interaction
parameters (\ref{eq:par}) is quite adequate, giving for the 
lowest state a mass of 5530 MeV, close to the observed value by the D\O~~Collaboration \cite{D0:2016mwd}, 
with a hyperfine 
contribution of  - 560 MeV.  Note that the state $\psi^1_{0^{+}}$ alone gives  - 512 MeV  
to this binding, as one can see from the first diagonal
matrix element of Eq. (\ref{0++}), namely $16(C_{13}+C_{24})$,  with the parameters 
of Eq. (\ref{eq:par}). This proves that the hyperfine contribution of the $u \bar d$ pair
is dominant in the system which means that one cannot neglect it. 

The hyperfine interaction is still attractive for  J$^{P}$ = 1$^{+}$ contributing with ~- 502 MeV.
Thus the lowest J$^{P}$ = 0$^{+}$ and J$^{P}$ = 1$^{+}$ are not degenerate, contrary to 
the diquark-antidiquark model \cite{Wang:2016tsi}.

However in the heavy quark limit, which can be simulated by taking $C_{i4}$ = 0, which follows 
from  $C_{i4} \propto 1/m_{heavy} \rightarrow 0 $ when $m_{heavy} \rightarrow \infty$ , one obtains 
%for the lowest states
%E(0$^{+}$) = 557 MeV and E(1$^{+}$) = 560 MeV, thus they become almost degenerate in this limit. 
degeneracy as follows.
The spectrum of 0$^{+}$ coincides with the first, second, third and sixth states of 1$^{+}$,
with eigenvalues 5573 MeV, 5851 MeV, 6133 MeV and 6319 MeV respectively
and the spectrum of 2$^{+}$  coincides with the fourth and fifth states of 1$^{+}$,
having as eigenvalues 6172 MeV and 6250 MeV respectively.

The hyperfine interaction is repulsive for J$^{P}$ = 2$^{+}$ states rising the pure mass term from 6090 MeV 
to 6183 MeV for the lowest state, as one can see from Fig. 2.

In the basis (\ref{eq:betai}) the lowest J$^{P}$ = 0$^{+}$ state 
has the amplitudes 
\begin{equation}
(-0.9193, 0.0246, -0.0852, 0.3833) 
\end{equation}
The first number implies that this state can decay substantially 
into a PP channel, \emph{i. e.} $B^0_s + \pi^{\pm}$ (threshold 5506 MeV).
The second number  indicates a negligible coupling to the  VV 
channel $B^{0*}_s + \rho^{\pm}$
which is consistent with the experiment.
The third and fourth numbers show the hidden-color content of the lowest state. 
The latter numbers should decrease asymptotically when the two quark-antiquark pairs separate.
As the  PP channel component of the lowest eigenstate is quite large it must be another 
reason to produce a narrow width, contrary to the case of X(3872) where a
tiny  VV channel component was a reason to explain its narrow width into 
$J/\psi +\rho$  or $J/\psi +\omega$ \cite{Hogaasen:2005jv}. The phase space 
could be an important factor.

Actually one can understand the reason why the two cases are so different. 
In the matrix produced by the chromomagnetic interaction for the $1^{++}$ 
case the key quantity was the off-diagonal matrix element 
$C_{23} - C_{12} \equiv C_{q \bar c} - C_{qc}$ = 1.5 MeV which is very small,
so there is no much coupling between the only color singlet-singlet and the only hidden color states. 
Then the amplitude of this VV channel is negligible and the hidden color part,
which does not decay, is dominant. Combined with the phase space of the decay
of X(3872) into $J/\psi +\rho$  or $J/\psi +\omega$ the lowest $1^{++}$ 
acquires a very small width. An analogue situation occurs for Y(4140) 
where $C_{23} - C_{12} \equiv C_{s \bar c} - C_{sc}$ = 1.7 MeV is also very small
\cite{Stancu:2009ka}.

In the present case there is no such tiny off-diagonal matrix element,
which one can infer from the definitions of these elements and the values of $C_{ij}$ 
given in Eq.(\ref{eq:par}).

On the other hand, there is a large phenomenological difference between X(3872) and 
X(5568). While  X(3872) is 
located only a few MeV below the 
$D \overline{D}^*$ threshold, the X(5568) resonance 
has a width about ten times larger and 
is quite far above the   $B^0_s + \pi^{\pm}$ threshold. This implies that X(5568) 
could be a more compact system, therefore a better candidate for tetraquarks.

Note however that the J$^{P}$ = 2$^{+}$ lowest state has more resemblance with the lowest $1^{++}$ 
state of  X(3872). Its amplitudes are 
\begin{equation}
(-0.1343, 0.9909) 
\end{equation}
which shows that the hidden color component is dominant and its decay width can be diminished 
in this way.

One can also introduce the exchange channel.
The corresponding
amplitudes can in principle  be obtained from the orthogonal transformation going from
the direct meson-meson  channel, Fig. 1b, 
to the exchange meson-meson  channel, Fig. 1c. 

Looking at Fig. 1c  and recalling that we chose  
1 = $u$, 2 = $s$, 3 = $\overline d$ and 4 = $\overline b$, for the 
exchange meson-meson channels we obtain
\begin{equation}
\psi^{1ex}_{0^{+}} = |1_{14}1_{23} \rangle |P_{14}P_{23} \rangle = 
B^+ \overline{K}^0
\end{equation}
\begin{equation}\label{B*K*}
\psi^{2ex}_{0^{+}} =  |1_{14}1_{23} \rangle |(V_{14}V_{23})_0 \rangle =
 B^{+*}  \overline{K}^{0*}
\end{equation}
The orthogonal transformation between the direct and exchange channel bases can be found in Appendix A 
of Ref. \cite{Stancu:2009ka}.
But the tetraquark state X(5568) can hardly decay into the exchange channel
$B^+ \overline{K}^{0}$, the threshold being too high, at 5777  MeV. 

Back to the lowest $1^{+}$  state, located at 5588 MeV, we can see that the channel  
$B^*_s$ +$\pi$ is kinematically allowed,   
being at 33 MeV above the threshold of 5555 MeV. 

Finally, we note that the $s u \bar d  \bar b$ system studied here forms 
together with  $s u \bar u  \bar b$ and $\frac{1}{\sqrt{2}}s (d \bar d - u \bar u) \bar b$ 
an isospin triplet, all members being degenerate in the present approach.  There is 
also an isosinglet partner  $\frac{1}{\sqrt{2 + \alpha^2}}s (u \bar u + d \bar d + \alpha~  s \bar s) \bar b$,
with  $\alpha$ for SU(3)-flavor breaking,
which would have a slightly larger mass because the $s \bar s$ subsystem introduces a less
attractive hyperfine contribution, the parameter  $C_{s \bar s} $ = 8.6  MeV \cite{Buccella:2006fn}
being smaller than  $C_{q \bar q} $ = 29.8 MeV
($q = u, d$) from Eq. (\ref{eq:par}). The isosinglet partner may  decay into
$B_s + \eta$ if the phase space allows.  Useful considerations about the decay and observation 
of the isoscalar partner were made in Ref. \cite{Ali:2016gdg}.

%%%%%%%%%%%%%%%%%%%%%%%%%%%%%%%%%%%%%%%%%%%%%%%%%%%%%%
\section{The spectrum of $s u \bar d  \bar c$}\label{sudc}

%%%%%%%%%%%%%%%%%%%%%%%%
% May 23, 2016
\begin{figure}\label{fig3}
%%%%%%%%%%%%%%%%%%%%%%%%%%%%%%%%%%%%%%%%%%%%%%%%%%%%%
%\thispagestyle{empty}
\definecolor{red}{rgb}{1,0,0} 
\begin{center}
\setlength{\unitlength}{0.75cm}
\begin{picture}(24,20)(0,0)
\def\level{\line(2,0){2}}
\thicklines
%%%%%%%%%%%%%%%%%%%%%%%%%%%%0+
\put(2.7,0){\large 0$^{+}$}
\put(1.5,2.128){\level}
\put(3.7,2.028){\large 2128}
 
\put(1.5,8.93){\level}
\put(3.7,8.83){\large 2468}
 
\put(1.5,15.97){\level}
\put(3.7,15.87){\large 2820}
 
\put(1.5,21.15){\level}
\put(3.7,21.05){\large 3079}
%%%%%%%%%%%%%%%%%%%%%%%%%%%%%1+
\put(6.5,0){\large  1$^{+}$}

\put(6.0,5.13){\level}
\put(8.2,4.93){\large 2278}

\put(6.0,10.53){\level}
\put(8.2,10.33){\large 2548}

\put(6.0,15.83){\level}
\put(8.2,15.63){\large 2813}

\put(6.0,16.53){\level}
\put(8.2,16.33){\large 2848}

\put(6.0,18.07){\level}
\put(8.2,17.87){\large 2925}

\put(6.0,19.63){\level}
\put(8.2,19.43){\large 3003}

%%%%%%%%%%%%%%%%%%%%%%%%%2++
\put(12.0,0){\large  2$^{+}$}

\put(11.0,17.49){\level}
\put(13.2,17.39){\large 2896}

\put(11.0,19.07){\level}
\put(13.2,18.97){\large 2975}
%%%%%%%%%%%%%%%%%%%%%%%%%
\end{picture}
\end{center}
\caption{The spectrum of the $s u \bar d  \bar c$ tetraquark. }
\end{figure}
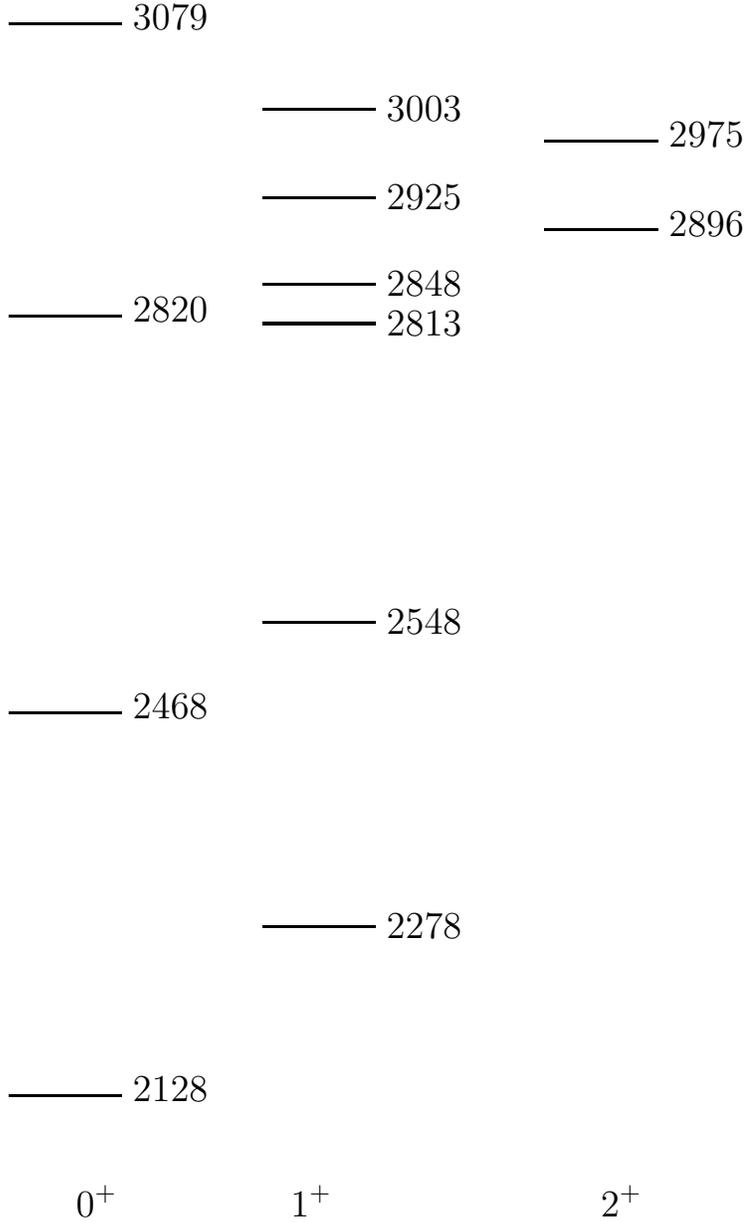

%%%%%%%%%%%%%%%%%%%%%%%%%%
We have also calculated the spectrum of the $s u \bar d  \bar c$ tetraquark using 
the matrices (\ref{0++}),  (\ref{1++}) and   (\ref{eq:tensor}). 
In addition to the parameters from row 1 Eq. (\ref{eq:par}),
we need $C_{q \bar c}$ coefficients. They are 
\begin{equation}\label{eq:parc}
\begin{array}{lll}
C_{u \bar c}=6.6\,\mathrm{MeV}, &C_{\bar d \bar c}=6.0\,\mathrm{MeV},
%C_{c \bar s}=2.1\,\mathrm{MeV}, 
&C_{s \bar c} = 6.7\,\mathrm{MeV}. 
\end{array}\end{equation}
where the first and the third were found in Ref.  \cite{Buccella:2006fn} and the second was identified 
with $C_{dc}$ of Ref. \cite{Hogaasen:2013nca}  where  $\Sigma_c - \Lambda_c$ and  $\Sigma^*_c -\Sigma_c$ splittings
were fitted.
%which fits the $J/\psi$ - $\eta_c$, 
%$D^* - D$ and $\Sigma_c - \Lambda_c$ and  $\Sigma^*_c -\Sigma_c$ splittings
%\cite{Hogaasen:2013nca}. 
We took $m_c$ = 1550 MeV \cite{Hogaasen:2005jv} like in our study on Y(4140) \cite{Stancu:2009ka}
and $m_{u,d}$ as in Eq. (\ref{eq:mass}).
This gives $\sum_i m_i$ = 2780 MeV. The lowest $0^+$ state has a mass of 2128 MeV 
from a hyperfine attraction of - 652 MeV and the lowest $1^+$ state has a mass of 2278MeV with  
a hyperfine attraction of - 502 MeV.  Thus there is a gap of 150 MeV between the  lowest $1^+$ state 
and the lowest $0^+$. 
The resulting masses for the lowest states are by about 200 MeV lower than in diquark-antidiquark studies 
of $s u \bar d  \bar c$ \cite{Wang:2016tsi}. But of course, the choice of quark masses imposes
some arbitrariness because they depend on the environment,
in particular, on the confinement. As pointed out in Ref. \cite{Buccella:2006fn} the values of 
$m_i$ extracted from baryons are usually higher than those from mesons. 

The pattern of the whole spectrum of $s u \bar d  \bar c$ and of
$s u \bar d  \bar b$ are very similar. The main difference is that the gap between
the  lowest $1^+$ state and the lowest $0^+$ state is about
twice as large as that of   $s u \bar d  \bar b$ system. This gap will
decrease in the heavy quark limit, making  $0^+$ and four of the $1^+$ states degenerate. The  $2^+$ state 
raises from 2780 MeV to 2896 MeV due to a repulsive hyperfine contribution of 116 MeV
and the highest  $1^+$ state is close to the highest $2^+$ state.
   
In this description the lowest $0^+$ state can decay into $D_s + \pi$ the channel for which the 
threshold is at 2108 MeV and  the lowest $1^+$ state can decay into $D^*_s + \pi$, the threshold being at
2252 MeV.

%%%%%%%%%%%%%%%%%%%%%%%%%%%%%%%%%%%%%%%%%%%%%%%%%%%%%%%%%%%%%%%%%%%%%%

\section{Conclusions}

We have presented results in a simple tetraquark model 
to see whether or not this model is compatible with the observation of the 
X(5568) resonance announced by the D\O~~Collaboration.
The parameters of the model were deduced from meson and
baryon masses. The calculated mass of X(5568) so obtained is in agreement 
with the experimental mass range found  by the D\O~~Collaboration. 
The large value of the chromomagnetic parameter $C_{u \bar d}$
and a complete color basis played key roles in obtaining a low mass for  X(5568). 
The structure of the lowest $0^+$ wave function   
indicates a large component in the $B_s^0\pi^+$ channel. Besides the $J^P$ = $0^+$
spectrum we also gave predictions for the  $J^P$ = $1^+$ and $2^+$ sectors.

In the critical analysis of various possible interpretations of X(5568)
(threshold, cusp, molecular, tetraquarks) 
performed in Ref. \cite{Burns:2016gvy} it has been argued that none of these interpretations
seems a natural fit for   X(5568). 
Although this resonance seems to be too light for a plausible tetraquark candidate,
the authors of Ref. \cite{Burns:2016gvy} consider that the present approach
seems most promising due to low quark masses and a large hyperfine contribution,
taken fully into account in a complete color basis.

A more elaborate study of the $s u  \bar d  \bar b$ tetraquark 
system is worth by itself. The tetraquarks containing four different flavors 
may be more difficult to study than those with one light and one heavy flavor.
It may be a long way before understanding them, if confirmed. The presence of 
light quarks or antiquarks 
challenges our understanding of QCD \cite{Guo:2016nhb}.

%%%%%%%%%%%%%%%%%%%%%%%%%%%%%%%%%%%%%%%%%%%%%%%%%%%%%%%%%%%%%%%%%%%%%%%%%%%

\vspace{1.5cm}
\noindent
\emph{Note added.}
The first version of this work was submitted to arXiv, prior to the  LHCb
announcement \cite{LHCb} described results only for the
 $J^P = 0^+$ states of the  $s u  \bar d  \bar b$ tetraquark. The present
version extends the study of the spectrum to  $J^P = 1^+$ and  $2^+$ as well 
in order to clarify 
a few issues raised in the literature meanwhile, and includes predictions for the charmed
partner.
%%%%%%%%%%%%%%%%%%%%%%%%%%%%%%%%%%%%%%%%%%%%%%%%%%%%%%%%%%%%%%%%%%%%%%%%%
\section{Acknowledgments} 

I am most grateful to Tim Burns for pointing out two incorrect matrix elements in
the matrix (10). This helped to clarify the issue on the degeneracy at infinite
mass limit of the heavy quark, but did not much change the spectrum of $1^+$ states.
This research was supported by the Fonds de la Recherche Scientifique - FNRS, Belgium,  
under the Grant No. 4.4501.05.
 
%%%%%%%%%%%%%%%%%%%%%%%%%%%%%%%%%%%%%%%%%%%%%%%%%%%%%%%%%%%%%%%%%%%%
\appendix

\section{Charge conjugation}\label{chargeconj}

From Ref. \cite{Stancu:1991rc} Ch. 10,  one
can see that the permutation (13)(24) leaves invariant the 
color basis vectors $|1_{13} 1_{24} \rangle$ and $|8_{13} 8_{24} \rangle $. 
Then, with the identification 1 = $u$, 2 = $s$, 3 = $\overline d$ and 
4 = $\overline b$ the permutation (13)(24) is equivalent to the
charge conjugation operator \cite{Stancu:2008zh}. Thus all basis states 
introduced in this way have a definite charge conjugation, which is easy to identify.

%%%%%%%%%%%%%%%%%%%%%%%%%%%%%%%%%%%%%%%%%%%%%%%%%%%%%%%%%%

%%%%%%%%%%%%%%%%%%%%%%%%%%%%%%%%%%%%%%%%%%%%%%%%%%%%%%%%%%%%%%%%%%%%%%%%

%%%%%%%%%%%%%%%%%%%%%%%%%%%%%%%%%%%%%%%%%%%%%%%%%%%%%%%%%%%%%%%%
\end{document}